\def\ref#1{\medskip\everypar={\hangindent 2\parindent}#1}
\def\beginref{\begingroup
\bigskip
\centerline{\bf References}
\nobreak\noindent}
\def\endref{\par\endgroup}

\def\PL#1{Phys.\ Lett.\ {\bf#1}}

\def\PR#1{Phys.\ Rev.\ {\bf#1}}\def\CQG#1{Class.\ Quantum Grav.\ {\bf#1}}

\def\JMP#1{J.\ Math.\ Phys.\ {\bf#1}}

\def\JHEP#1{JHEP\ {\bf#1}}\def\JCAP#1{JCAP\ {\bf#1}}\def\EPJ#1{Eur.\ Phys.\ J.\ {\bf#1}}
\def\RMP#1{Rev.\ Mod.\ Phys.\ {\bf#1}}

\def\arx#1{{\tt arXiv:#1}}

\def\hx{\hat x}\def\hp{\hat p}

\def\lra{\leftrightarrow}
\def\ha{{1\over2}}\def\qu{{1\over4}}
\def\section#1{\bigskip\noindent{\bf#1}\smallskip}

\def\dg{\dagger}\def\sqx{\sqrt{1-\alpha^2x^2}}\def\sqp{\sqrt{1-\beta^2p^2}}\def\sqk{\sqrt{1-\alpha^2k^2}}
\def\gtt{g_{22}}\def\gtf{g_{24}}\def\gft{g_{42}}\def\nm{{\nu\mu}}\def\mn{{\mu\nu}}
\def\act{\triangleright}\def\ad{{\rm ad}}

{\nopagenumbers
\line{}
\vskip40pt
\centerline{\bf Hermitian realizations of the Yang model}
\vskip50pt
\centerline{{\bf T. Martini\'c-Bila\'c}\footnote{$^\ddagger$}{e-mail: teamar@pmfst.hr}}
\vskip5pt
\centerline {Faculty of Science, University of Split,}
\centerline{Rudjera Bo\v skovi\'ca 33, 21000 Split, Croatia}
\vskip10pt
\centerline{{\bf S. Meljanac}\footnote{$^\dagger$}{e-mail: meljanac@irb.hr}}
\vskip5pt
\centerline {Rudjer Bo\v skovi\'c Institute, Theoretical Physics Division}
\centerline{Bljeni\v cka c.~54, 10002 Zagreb, Croatia}
\vskip10pt
\centerline{{\bf S. Mignemi}\footnote{$^\ast$}{e-mail: smignemi@unica.it}}
\vskip5pt
\centerline {Dipartimento di Matematica, Universit\`a di Cagliari}
\centerline{via Ospedale 72, 09124 Cagliari, Italy}
\smallskip
\centerline{and INFN, Sezione di Cagliari}
\centerline{Cittadella Universitaria, 09042 Monserrato, Italy}

\vskip40pt
\centerline{\bf Abstract}
\medskip
{\noindent The Yang model is an example of noncommutative geometry on a background spacetime of constant curvature.
We discuss the Hermitian realizations of its associated algebra on phase space in a perturbative expansion up to sixth order.
We also discuss its realizations on extended phase spaces, that include additional tensorial and/or vectorial
degrees or freedom.}
\vskip60pt
\vfil\eject}

\section{1. Introduction}
In recent years noncommutative models in curved spacetime have been extensively investigated, either from a formal
perspective [1-12], also in connection with quantum field theory [13-15],
or in view of their application in the study of phenomenological effects in cosmology [16-17].

However, the first example of noncommutativity on a curved spacetime background was proposed by C.N. Yang [18]
already in 1947, soon after Snyder had introduced the idea of a noncommutative spacetime [19].
Yang's proposal was based on an algebra which included phase space and Lorentz
generators, where the commutation relations  between the components of the position operators, as well as those of the momentum operators
were not trivial, giving rise to a spacetime displaying both noncommutativity and curvature.

The noncommutative Yang algebra is a 15-parameter algebra, isomorphic to $so(1,5)$, defined by the relations
$$\eqalignno{&[\hx_\mu,\hx_\nu]=i\beta^2M_\mn,\qquad[\hp_\mu,\hp_\nu]=i\alpha^2M_\mn,\qquad[\hx_\mu,\hp_\nu]=i\eta_\mn h,\cr
&[h,\hx_\mu]=i\beta^2\hp_\mu,\qquad[h,\hp_\mu]=-i\alpha^2\hx_\mu,\qquad[M_\mn,h]=0,\cr
&[M_\mn,\hx_\lambda]=i(\eta_{\mu\lambda}\hx_\nu-\eta_{\nu\lambda}\hx_\mu),\qquad[M_\mn,\hp_\lambda]=i(\eta_{\mu\lambda}\hp_\nu-\eta_{\nu\lambda}\hp_\mu),\cr
&[M_\mn,M_{\rho\sigma}]=i\big(\eta_{\mu\rho}M_{\nu\sigma}-\eta_{\mu\sigma}M_{\nu\rho}-\eta_{\nu\rho}M_{\mu\sigma}+\eta_{\nu\sigma}M_{\mu\rho}\big),&(1)}$$
where $\alpha$ and $\beta$ are real parameters and $\eta_\mn$ the flat metric and we use natural units, $\hbar=c=1$.

We interpret the operators $\hx_\mu$ and $\hp_\mu$ as coordinates of the quantum phase space, $M_\mn$ as generators of
the Lorentz transformations and $h$ as a further scalar generator, necessary to close the algebra.
The algebra (1) is invariant under Born duality [20], $\alpha\lra\beta$, $\hx_\mu\to-\hp_\mu$,  $\hp_\mu\to\hx_\mu$, $M_\mn\lra M_\mn$,
$h\lra h$.
It contains as subalgebras both the de Sitter and the Snyder algebras, to which it reduces in the limit $\beta\to0$ and
$\alpha\to0$, respectively.

We have investigated the Yang model in previous papers. In particular, in [10-11] we have considered noncommutative models
in a spacetime of constant curvature and discussed their realizations on a quantum phase space. These models preserve the
Lorentz invariance and, besides Yang proposal, include some generalizations [1-2].
Later, in [12], we have discussed the possibility of obtaining Yang model by symmetry breaking of an algebra
defined in an extended quantum phase space that includes also tensorial generators, of the kind introduced in [21-25]
for the Snyder model.

Following [12],
in this paper we shall investigate the realizations of the Yang algebra in terms of a restricted number of operators of a Hilbert
space, the simplest case being realizations in terms of phase space variables $x_\mu$ and $p_\mu$ [10-11]. However, we shall use
a more efficient procedure than in [12] for going to higher orders. Moreover,
several possibilities arise depending on how many operators are introduced to generate the Hilbert space, as we discuss in the
following sections. For example, one may consider the Lorentz generators as independent from the phase space ones, as proposed
in [21-24] in the case of the Snyder model.
Some choices may be useful to obtain Hopf algebra structures, which are not possible in a phase space realization.
In general, we shall only obtain perturbative realizations of the algebra, since analytic results seem to be out of reach.

\section{2. Realizations of Yang model on quantum phase space}
In this section, we look for Hermitian realizations in quantum phase space, with
$$\hx_\mu^\dg=\hx_\mu,\quad\hp_\mu^\dg=\hp_\mu,\quad M_\mn^\dg=M_\mn,\quad h^\dg=h,\eqno(2)$$
where $\hx_\mu$, $\hp_\mu$, $M_\mn$ and $h$ are functions of phase space operators $x_\mu$ and $p_\mu$ that satisfy the
Heisenberg algebra
$$[x_\mu,x_\nu]=[p_\mu,p_\nu]=0,\qquad[x_\mu,p_\nu]=i\eta_\mn,\eqno(3)$$
with
$$M_\mn=x_\mu p_\nu-x_\nu p_\mu,\eqno(4)$$
and $p_\mu\act1=0$, $M_\mn\act1=0$.

In the limit $\alpha=0$, a Hermitian realization of the algebra (1) is given by [11]
$$\hx_\mu(\beta)=\ha\left(x_\mu\sqp+\sqp\,x_\mu\right),\qquad\hp_\mu=p_\mu,\qquad h=\sqp.\eqno(5)$$
Analogously, when $\beta=0$, a realization is
$$\hp_\mu(\alpha)=\ha\left(p_\mu\sqx+\sqx\,p_\mu\right),\qquad\hx_\mu=x_\mu,\qquad h=\sqx.\eqno(6)$$

However, when both $\alpha\ne0$ and $\beta\ne0$, we get
$$\eqalignno{&[\hx_\mu(\beta),\hp_\nu(\alpha)]=\cr
&\quad{i\over2}\,\eta_\mn\left(\sqrt{(1-\alpha^2x^2)(1-\beta^2p^2)}+\sqrt{(1-\beta^2p^2)(1-\alpha^2x^2)}\,\right)+\qu x_\mu(p_\nu K+ Kp_\nu)+\qu(p_\nu K+ Kp_\nu)x_\mu=\quad\cr
&\quad{i\over2}\,\eta_\mn\left(\sqrt{(1-\alpha^2x^2)(1-\beta^2p^2)}+\sqrt{(1-\beta^2p^2)(1-\alpha^2x^2)}\,\right)+\qu p_\nu(x_\mu K+ Kx_\mu)+\qu(x_\mu K+ Kx_\mu)p_\nu,\quad&(7)}$$
where
$$\eqalignno{&K=\sum_{m,n=0}^\infty{\ha\choose m}{\ha\choose n}(-\beta^2)^m(-\alpha^2)^n[p^{2m},x^{2n}]=\cr
&\quad-i\left(\alpha^2\beta^2D+{\alpha^2\beta^4\over4}(p^2D+Dp^2)+{\alpha^4\beta^2\over4}(x^2D+Dx^2)+\dots\right),&(8)}$$
with $D=\ha(x\cdot p+p\cdot x)$.\footnote{$^1$}{We denote $x^2=x_\alpha x_\alpha$, $x\cdot p=x_\alpha p_\alpha$, and so on.}

This result is different from $i\eta_\mn h(x,p)$ and therefore $\hx_\mu(\beta)$ and $\hp_\mu(\alpha)$ are not a realization of the Yang algebra.
In order to construct a true realization of the Yang algebra, we fix $\hp_\mu=\hp_\mu(\alpha)$ and define $\hx_\mu=e^{iG}\hx_\mu(\beta)e^{-iG}$,
choosing $G$ such that $[\hx_\mu,\hp_\nu]=i\eta_\mn h$. In general, we can expand $G$ as
$$G=\sum_{m,n=1}^\infty\alpha^{2m}\beta^{2n}g_{2m,2n},\eqno(9)$$
where $g_{2m,2n}$ are functions of $x^2$, $p^2$ and $D$.
From $\hx_\mu=e^{iG}\hx_\mu(\beta)e^{-iG}$, it follows
$$\hx_\mu=\hx_\mu(\beta)+i[G,\hx_\mu(\beta)]+{i^2\over2!}\big[G,[G,\hx_\mu(\beta)]\big]+\dots\eqno(10)$$
Then, up to sixth order in $\alpha$ and $\beta$, we get
$$[G,\hx_\mu(\beta)]=\alpha^2\beta^2[\gtt,x_\mu]-{\alpha^2\beta^4\over4}[\gtt,x_\mu p^2+p^2x_\mu]+\alpha^2\beta^4[\gtf,x_\mu]+\alpha^4\beta^2[\gft,x_\mu].\eqno(11)$$

Hence,
$$\eqalignno{&[\hx_\mu,\hp_\nu]=[\hx_\mu(\beta),\hp_\nu(\alpha)]+i\alpha^2\beta^2[[\gtt,x_\mu],p_\nu]-i{\alpha^2\beta^4\over4}[[\gtt,x_\mu p^2+p^2x_\mu],p_\nu]\cr
&\quad+i\alpha^2\beta^4[[\gtf,x_\mu],p_\nu]+i\alpha^4\beta^2[[\gft,x_\mu],p_\nu]-{i\over4}\alpha^4\beta^2[[\gtt,x_\mu],p_\nu x^2+x^2p_\nu].&(12)}$$
Substituting in (7), it follows
$$\eqalignno{&[\hx_\mu,\hp_\nu]=\cr
&\quad{i\over2}\eta_\mn\left(\sqrt{(1-\alpha^2x^2)(1-\beta^2p^2)}+\sqrt{(1-\beta^2p^2)(1-\alpha^2x^2)}\,\right)-{i\alpha^2\beta^2\over4}\Big(x_\mu(p_\nu D+ Dp_\nu)+(p_\nu D+Dp_\nu)x_\mu\Big)\cr
&\quad-{i\alpha^2\beta^4\over16}\bigg(x_\mu\Big(p_\nu(p^2D+Dp^2)+(p^2D+Dp^2)p_\nu\Big)+\Big(p_\nu(p^2D+Dp^2)+(p^2D+Dp^2)p_\nu\Big)x_\mu\bigg)\cr
&\quad-{i\alpha^4\beta^2\over16}\bigg(x_\mu\Big(p_\nu(x^2D+Dx^2)+(x^2D+Dx^2)p_\nu\Big)+\Big(p_\nu(x^2D+Dx^2)+(x^2D+Dx^2)p_\nu\Big)x_\mu\bigg)\cr
&\quad+i\alpha^2\beta^2\big[\big[\gtt,x_\mu\big],p_\nu\big]-i{\alpha^2\beta^4\over4}\big[\big[\gtt,x_\mu p^2+p^2x_\mu\big],p_\nu\big]
+i\alpha^2\beta^4\big[\big[\gtf,x_\mu\big],p_\nu\big]+i\alpha^4\beta^2\big[\big[\gft,x_\mu\big],p_\nu\big]\cr
&\quad-{i\over4}\alpha^4\beta^2\big[\big[\gtt,x_\mu\big],p_\nu x^2+x^2p_\nu\big].&(13)}$$

Requiring that only terms proportional to $\eta_\mn$ survive, one obtains (see Appendix A)
$$\gtt={1\over6}\left(D^3-\ha D\right),\qquad\gtf=-{1\over16}(Dp^2+p^2D),\qquad\gft=-{1\over16}(Dx^2+x^2D).\eqno(14)$$
Hence, at this order,
$$G={\alpha^2\beta^2\over6}\left(D^3-\ha D\right)-{\alpha^2\beta^2\over16}\Big(D(\alpha^2x^2+\beta^2p^2)+(\alpha^2x^2+\beta^2p^2)D\Big),\eqno(15)$$
and then
$$\eqalignno{\hx_\mu&={1\over2}\left(x_\mu\sqrt{(1-\beta^2p^2)}+\sqrt{(1-\beta^2p^2)}\,x_\mu\right)+{\alpha^2\beta^2\over4}\Big(x_\mu D^2+D^2x_\mu\Big)\cr
&-{\alpha^2\beta^2\over16}\bigg[x_\mu(\alpha^2x^2+\beta^2p^2)+(\alpha^2x^2+\beta^2p^2)x_\mu+2\beta^2(Dp_\mu+p_\mu D)\cr
&-\beta^2\Big((x_\mu p^2+p^2x_\mu)D^2+D^2(x_\mu p^2+p^2x_\mu)\Big)\bigg],\cr
\hp_\mu&=\ha\left(p_\mu\sqx+\sqx\,p_\mu\right),&(16)}$$
and $[\hx_\mu,\hp_\nu]=i\eta_\mn h$, with
$$\eqalignno{h&=\ha\left(\sqrt{(1-\alpha^2x^2)(1-\beta^2p^2)}+\sqrt{(1-\beta^2p^2)(1-\alpha^2x^2)}+\alpha^2\beta^2D^2\right)\cr
&+{\alpha^2\beta^2\over8}\bigg(\beta^2p^2-\alpha^2x^2+\beta^2(p^2D^2+D^2p^2)-\alpha^2(x^2D^2+D^2x^2)\bigg).&(17)}$$

We point out that infinitely many realizations can be obtained from $\hx_\mu$ and $\hp_\mu$ in (16) by similarity transformations,
defined by acting simultaneously with $e^{iG(D\,,x^2,p^2)}$ on all generators $\hx_\mu$, $\hp_\mu$, $M_\mn$ and $h$ obtained above.
Note that $M_\mn$ is invariant under these transformations because $G$ is a function of Lorentz-invariant operators, but $h$ is
not invariant since $[G,h]\ne 0$.

Special classes of realizations are
$$\eqalignno{&\hx_\mu(c_1)=e^{-ic_2G}\hx_\mu e^{ic_2G}=e^{ic_1G}\hx_\mu(\beta)\,e^{-ic_1G}\cr
&\hp_\mu(c_2)=e^{-ic_2G}\hp_\mu e^{ic_2G}=e^{-ic_2G}\hp_\mu(\alpha)\,e^{ic_2G}\cr
&h(c_1,c_2)=e^{-ic_2G}h\,e^{ic_2G}&(18)}$$
with $G$ given in (15) and $c_1+c_2=1$.

At sixth order in $\alpha$ in $\beta$ we have
$$\eqalignno{\hx_\mu(c_1)&={1\over2}\left(x_\mu\sqrt{(1-\beta^2p^2)}+\sqrt{(1-\beta^2p^2)}\,x_\mu\right)+c_1{\alpha^2\beta^2\over4}\Big(x_\mu D^2+D^2x_\mu\Big)\cr
&-c_1{\alpha^2\beta^2\over16}\bigg[x_\mu(\alpha^2x^2+\beta^2p^2)+(\alpha^2x^2+\beta^2p^2)x_\mu+2\beta^2(Dp_\mu+p_\mu D)\cr
&-\beta^2\Big((x_\mu p^2+p^2x_\mu)D^2+D^2(x_\mu p^2+p^2x_\mu)\Big)\bigg]\cr
&\cr
\hp_\mu(c_2)&={1\over2}\left(p_\mu\sqrt{(1-\alpha^2x^2)}+\sqrt{(1-\alpha^2x^2)}\,p_\mu\right)+c_2{\alpha^2\beta^2\over4}\Big(p_\mu D^2+D^2p_\mu\Big)\cr
&-c_2{\alpha^2\beta^2\over16}\bigg[p_\mu(\alpha^2x^2+\beta^2p^2)+(\alpha^2x^2+\beta^2p^2)p_\mu+2\alpha^2(Dx_\mu+x_\mu D)\cr
&-\alpha^2\Big((p_\mu x^2+x^2p_\mu)D^2+D^2(p_\mu x^2+x^2p_\mu)\Big)\bigg]\cr
&\cr
h(c_1,c_2)&=\ha\left(\sqrt{(1-\alpha^2x^2)(1-\beta^2p^2)}+\sqrt{(1-\beta^2p^2)(1-\alpha^2x^2)}+\alpha^2\beta^2D^2\right)\cr
&+(c_1-c_2){\alpha^2\beta^2\over8}\bigg(\beta^2p^2-\alpha^2x^2+\beta^2(p^2D^2+D^2p^2)-\alpha^2(x^2D^2+D^2x^2)\bigg).&(19)}$$

In particular, for $c_1=c_2=\ha$,
$$h\left(\ha,\ha\right)=\ha\left(\sqrt{(1-\alpha^2x^2)(1-\beta^2p^2)}+\sqrt{(1-\beta^2p^2)(1-\alpha^2x^2)}+\alpha^2\beta^2D^2\right),\eqno(20)$$
and
$$[h(c_1,c_2),\hx_\mu(c_1)]=i\beta^2\hp_\mu(c_2),\qquad[h(c_1,c_2),\hp_\mu(c_2)]=-i\alpha^2\hx_\mu(c_1).\eqno(21)$$

\section{3. Realizations of extended Yang model on quantum phase space}
A different realization of the Yang algebra can be obtained introducing additional tensorial generators $\hx_\mn=-\hx_\nm$,
similarly to what has been done in [21-25] for the Snyder model or in [26] for a more general setting. They are assumed to satisfy
$$\eqalignno{&[\hx_\mn,\hx_{\rho\sigma}]=i\big(\eta_{\mu\rho}\hx_{\nu\sigma}-\eta_{\mu\sigma}\hx_{\nu\rho}-\eta_{\nu\rho}\hx_{\mu\sigma}+\eta_{\nu\sigma}\hx_{\mu\rho}\big),\cr
&[\hx_\mn,x_\lambda]=0,\qquad\qquad[\hx_\mn,p_\lambda]=0.&(22)}$$
In this case, we consider realizations of Lorentz generators of the form
$$M_\mn=\hx_\mn+x_\mu p_\nu-x_\nu p_\mu,\eqno(23)$$
and $M_\mn\act1=\hx_\mn\act1=x_\mn$, where $x_\mn$ are commuting variables.

In the limit $\alpha=0$, a realization of the Yang algebra is given by
$$\hx_\mu(\beta)=\ha\left(x_\mu\sqp+\sqp x_\mu\right)-\beta^2\hx_{\mu\alpha}\,{p_\alpha\over 1+\sqp},\qquad\hp_\mu=p_\mu,\qquad h=\sqp.\eqno(24)$$
Analogously, when $\beta=0$,
$$\hp_\mu(\alpha)=\ha\left(p_\mu\sqx+\sqx p_\mu\right)+\alpha^2\hx_{\mu\alpha}\,{x_\alpha\over 1+\sqx},\qquad\hx_\mu=x_\mu,\qquad h=\sqx.\eqno(25)$$

Also in this case, if both $\alpha\ne0$ and $\beta\ne0$, $\hx_\mu(\beta)$ and $\hp_\mu(\alpha)$ do not constitute a realization of the Yang algebra, since
$[\hx_\mu(\beta),\hp_\nu(\alpha)]\ne i\eta_\mn h$. Therefore,
as in the previous section, in order to construct a realization in terms of the extended algebra (22),
we fix $\hp_\mu=\hp_\mu(\alpha)$
and define $\hx_\mu=e^{iG}\hx_\mu(\beta)e^{-iG}$, constructing the operator $G$ in such a way that $[\hx_\mu,\hp_\nu]=i\eta_\mn h$.

From the expansion (9), we get at fourth order in $\alpha$, $\beta$,
$$G=\alpha^2\beta^2\bigg[{1\over6}\Big(D^3-\ha D\Big)-{1\over8}\hx_{\alpha\beta}(x_\alpha p_\beta+p_\beta x_\alpha)D-
{1\over8}\hx_{\alpha\gamma}\hx_{\beta\gamma}(x_\alpha p_\beta+p_\alpha x_\beta)\bigg].\eqno(26)$$
Hence,
$$\eqalignno{&\hx_\mu=\hx_\mu(\beta)+{\alpha^2\beta^2\over4}\big(x_\mu D^2+D^2x_\mu\big)+{\alpha^2\beta^2\over4}\hx_{\mu\alpha}x_\alpha D
-{\alpha^2\beta^2\over8}\Big(\hx_{\alpha\beta}(x_\alpha p_\beta+p_\beta x_\alpha)x_\mu\Big)\cr
&\quad-{\alpha^2\beta^2\over8}\big(\hx_{\alpha\gamma}\hx_{\mu\gamma}+\hx_{\mu\gamma}\hx_{\alpha\gamma}\big)x_\alpha,&(27)}$$
and
$$h=\ha\left(\sqrt{(1-\alpha^2x^2)(1-\beta^2p^2)}+\sqrt{(1-\beta^2p^2)(1-\alpha^2x^2)}+\alpha^2\beta^2D^2\right).\eqno(28)$$

There are infinitely many realizations obtained from $\hx_\mu$ and $\hp_\mu$ with arbitrary similarity transformations
that are invariant under Lorentz transformations and act on all generators $\hx_\mu$, $\hp_\mu$ and $h$ simultaneously. Note
that $M_\mn$ is invariant under these transformations but $h$ is not.

\section{4. Realizations of Yang model on double quantum phase space}
A different class of realizations can be obtained by adding to the generators $x_\mu$, $p_\mu$ of the Heisenberg algebra
new generators $q_\mu$ and $k_\mu$ satisfying a second Heisenberg algebra,
$$[q_\mu,q_\nu]=0,\qquad[k_\mu,k_\nu]=0,\qquad[q_\mu,k_\nu]=i\eta_\mn,\eqno(29)$$
with
$$[x_\mu,q_\nu]=[x_\mu,k_\nu]=0,\qquad[p_\mu,q_\nu]=[p_\mu,k_\nu]=0,\eqno(30)$$
and
$$p_\mu\act1=0,\quad k_\mu\act1=0,\quad x_\mu\act1=x_\mu,\quad q_\mu\act1=q_\mu.\eqno(31)$$
These realizations are more symmetric in the phase space variables and might permit the definition of a Hopf structure.
We shall call the phase space obtained by the addition of  $q_\mu$ and $k_\mu$ double quantum phase space.

In the limit $\alpha\to0$, a realization of the Yang model in this space is given by
$$\eqalignno{&\hx_\mu(\beta)=\ha\left(x_\mu\sqp+\sqp\,x_\mu\right)+{b\over2}\left(k_\mu\sqrt{1-{\beta^2q^2\over b^2}}+\sqrt{1-{\beta^2q^2\over b^2}}\,k_\mu\right),\cr
&\hp_\mu=q_\mu+\tilde b\,p_\mu,\qquad h=\tilde b\sqp-b\sqrt{1-{\beta^2q^2\over b^2}},\qquad M_\mn=x_\mu p_\nu-x_\nu p_\mu+q_\mu k_\nu-q_\nu k_\mu,&(32)}$$
with nonvanishing parameters $b$ and $\tilde b$, with $\tilde b-b=1$.
Analogously, when $\beta=0$,
$$\eqalignno{&\hp_\mu(\alpha)=\ha\left(q_\mu\sqk+\sqk\,q_\mu\right)+{\tilde b\over2}\left(p_\mu\sqrt{1-{\alpha^2x^2\over\tilde b^2}}+\sqrt{1-{\alpha^2x^2\over\tilde b^2}}\,p_\mu\right),\cr
&\hx_\mu=x_\mu+b\,k_\mu,\qquad h=-b\sqk+\tilde b\sqrt{1-{\alpha^2x^2\over\tilde b^2}},\qquad M_\mn=x_\mu p_\nu-x_\nu p_\mu+q_\mu k_\nu-q_\nu k_\mu.&(33)}$$
As usual, if both $\alpha\ne0$ and $\beta\ne0$, $\hx_\mu(\beta)$ and $\hp_\mu(\alpha)$ are not a realization of the Yang algebra.

In order to construct realizations of the Yang model in this space, as in sections 2 and 3, we set
$\hp_\mu=\hp_\mu(\alpha)$ and $\hx_\mu=e^{iG}\hx_\mu(\beta)e^{-iG}$, and construct the operator $G$ such that $[\hx_\mu,\hp_\nu]=i\eta_\mn h$.
In general, $G$ can be expanded as in (9).

Proceeding as usual, we get at fourth order in $\alpha$ and $\beta$,
$$G={\alpha^2\beta^2\over6}\left[{1\over\tilde b^2}\left(D^3-\ha D\right)-{1\over b^2}\left(\tilde D^3-\ha\tilde D\right)\right],\eqno(34)$$
where $\tilde D=\ha (k\cdot q+q\cdot k)$.
Hence,
$$\hx_\mu=\hx_\mu(\beta)+{\alpha^2\beta^2\over4}\left[{1\over\tilde b^2}(x_\mu D^2+D^2x_\mu)+{1\over b}(k_\mu\tilde D^2+\tilde D^2k_\mu)\right],\eqno(35)$$
and
$$\eqalignno{&h={\tilde b\over2}\left(\sqrt{(1-\beta^2p^2)\left(1-{\alpha^2x^2\over\tilde b^2}\right)}+\sqrt{\left(1-{\alpha^2x^2\over\tilde b^2}\right)(1-\beta^2p^2)}\,\right)\cr
&+{b\over2}\left(\sqrt{(1-\alpha^2k^2)\left(1-{\beta^2q^2\over b}\right)}+\sqrt{\left(1-{\beta^2q^2\over b}\right)(1-\alpha^2k^2)}\,\right)
+{\alpha^2\beta^2\over2}\left({1\over\tilde b}D^2-{1\over b}\tilde D^2\right)&(36)}$$

Again, infinitely many realizations can be obtained by acting on (35), (36) with similarity transformations.

\section{5. Realizations of extended Yang model on double quantum phase space}

Let us finally consider the Yang model with both additional phase space generators and additional Lorentz generators.
Realizations of this kind have been considered in [12] in a slightly different formalism, see sect.~6.
The additional Lorentz generators $\hx_\mn$ are introduced as in sect.~3, such that
$$M_\mn=\hx_\mn+x_\mu p_\nu-x_\nu p_\mu+q_\mu k_\nu-q_\nu k_\mu,\eqno(37)$$
and $M_\mn\act1=\hx_\mn\act1=x_\mn$, where $x_\mn$ are commutative parameters.

Proceeding as usual, one can show that realizations up to second order in $\alpha^2$, $\beta^2$ are in this case
$$\eqalignno{&\hx_\mu=x_\mu-{\beta^2\over4}(x_\mu p^2+p^2x_\mu)+b\,k_\mu-{\beta^2\over4b}(k_\mu q^2+q^2k_\mu)-{\beta^2\over2\tilde b}(x_{\mu\alpha}p_\alpha-x_{\mu\alpha}q_\alpha),\cr
&\hp_\mu=q_\mu-{\alpha^2\over4}(q_\mu k^2+k^2q_\mu)+\tilde b\,p_\mu-{\alpha^2\over4\tilde b}(p_\mu x^2+x^2p_\mu)+{\alpha^2\over2b}(x_{\mu\alpha}k_\alpha+x_{\mu\alpha}x_\alpha),&(38)}$$
with
$$h=1-\ha\left({\alpha^2\over\tilde b}x^2+\beta^2\tilde b\,p^2-{\beta^2\over b}q^2-\alpha^2b\,k^2\right).\eqno(39)$$
Also in the present case infinitely many realizations can be obtained by similarity transformations.

\section{6. Concluding remarks}
In this paper we have  assumed that
in the limit $\alpha=0$, $\beta=0$, the Yang algebra (1) reduces to the ordinary Heisenberg algebra with Lorentz algebra action and $h\to 1$.
Realizations are obtained in terms of quantum phase space and double quantum phase space with or without tensorial coordinates.

An approach to the Yang algebra alternative to the one we have considered here is to view it as a Lie algebra with 15 generators $\hx_\mu$, $\hp_\mu$,
$M_\mn$ and $\hat h$. When all structure constants go to zero, it reduces to a commutative space with coordinates $x_\mu$, $q_\mu$, $x_\mn$ and $h$
with relations $\hx_\mu\act1=x_\mu$, $\hp_\mu\act1=q_\mu$, $M_\mn\act1=x_\mn$ and $\hbar=0$.
Realizations of this Yang algebra can be found using the method of realizations of Lie algebras described in [9,26].
These realizations are linear in the position coordinates, but are given by power series in the momenta.
Such approach was used in [22-24] for the extended Snyder model and in [12] for the Yang model.

Finally, let us notice that the Yang model can be obtained from the $so(1,5)$ algebra with 15 generators $M_{AB}$ ($A,B=1,\dots5$), through
the relations $\hx_\mu=\beta M_{\mu4}$, $\hp_\mu=\alpha M_{\mu5}$ and $\hat h=\alpha\beta M_{45}$. A realization of $so(1,5)$ in symmetric ordering has been
presented in [26] and can be used for the Yang model as well.

As future prospects of our investigations we may envisage the possibility of constructing a star product and a twist using the double quantum
phase space.
Also the definition of a field theory on a spacetime based on the Yang model can be pursued from the present results and would be of great interest.

\section{Ackowledgements}
S. Mignemi acknowledges support from GNFM and COST action CA18108.

\bigskip
\beginref
\ref [1] V.V. Khruschev and A.N. Leznov, Grav. Cosmol. 9, 159 (2003).
\ref [2] J. Kowalski-Glikman and L. Smolin, \PR{D70}, 065020 (2004).
\ref [3] H.G. Guo, C.G. Huang and H.T. Wu, \PL{B663}, 270 (2008).
\ref [4] S. Mignemi, \CQG{26}, 245020 (2009).
\ref [5] M.C. Carrisi and S. Mignemi, \PR{D82}, 105031 (2010).
\ref [6] R. Banerjee, K. Kumar and D. Roychowdhury, \JHEP{1103}, 060 (2011).
\ref [7] A. Ballesteros, G. Gubitosi and F. Mercati, Symmetry {\bf 13}, 2099 (2021).
\ref [8] J. Lukierski and M. Woronowicz, \PL{B824}, 136783 (2021).
\ref [9] S. Meljanac and R. \v Strajn, SIGMA {\bf18}, 022 (2022).
\ref [10] S. Meljanac and S. Mignemi, \PL{B833}, 137289 (2022).
\ref [11] S. Meljanac and S. Mignemi, \JMP{64}, 023505 (2023).
\ref [12] J. Lukierski, S. Meljanac, S. Mignemi and A. Pachol, \arx{2212.02316}.
\ref [13] A. Schenkel and C.F. Uhlemann, SIGMA{\bf 6}, 061 (2010).
\ref [14] M. Buri\'c and M. Wohlgennant, \JHEP{1003}, 053 (2010).
\ref [15] A. Franchino-Vi\~nas and S. Mignemi, \EPJ{C80}, 382 (2020).
\ref [16] G. Rosati, G. Amelino-Camelia, A. Marciano and M. Matassa, \PR{D92}, 124042 (2015).
\ref [17] P. Aschieri, A. Borowiec and A. Pachol, \JCAP{04}, 025 (2021).
\ref [18] C.N. Yang, \PR{72}, 874 (1947).
\ref [19] H.S. Snyder, \PR{71}, 38 (1947).
\ref [20] M. Born, \RMP{21}, 463 (1949).
\ref [21] F. Girelli and E. Livine, \JHEP{1103}, 132 (2011).
\ref [22] S. Meljanac and S. Mignemi, \PR{D102}, 126011 (2020).
\ref [23] S. Meljanac and S. Mignemi, \PL{B814}, 136117 (2021).
\ref [24] S. Meljanac and S. Mignemi, \PR{D104}, 086006 (2021).
\ref [25] J. Lukierski, S. Meljanac, S. Mignemi and A. Pachol, \PL{B838}, 137709 (2023).
\ref [26] S. Meljanac, T. Martini\'c-Bila\'c and S. Kre\v sic-Juri\'c, \JMP{61}, 051705 (2020).

\endref

\bigbreak
\section{Appendix A}
In this appendix we give some details on the calculations leading to (14).

Starting from (13), we can compute the terms proportional to $\alpha^2\beta^2$.
These are given by
$$\eqalignno{&i\Big[[g_{22},x_\mu],p_\nu\Big]-{i\over4}\Big(x_\mu(p_\nu D+Dp_\nu)+(p_\nu D+Dp_\nu)x_\mu\Big)=\cr
&\quad\Big[i[g_{22},x_\mu]-\qu(x_\mu D ^2+D^2x_\mu),p_\nu\Big]+{i\over2}\eta_\mn D^2.&(A.1)}$$
Requiring that only terms proportional to $\eta_\mn$ survive in (A.1), one gets
$$i[g_{22},x_\mu]=\qu(x_\mu D ^2+D^2x_\mu),\eqno(A.2)$$
which is solved by
$$\gtt={1\over6}\left(D^3-\ha D\right).\eqno(A.3)$$

Then,
$$i\bigg[\Big[\gtt,\ha(x_\mu p^2+p^2x_\mu)\Big],p_\nu\bigg]\approx-{1\over8}\Big((x_\mu p^2+p^2x_\mu)A_\nu+A_\nu(x_\mu p^2+p^2x_\mu)\Big),\eqno(A.4)$$
and
$$i\left[[\gtt,x_\mu],\ha(p_\nu x^2+x^2p_\nu)\right]\approx-{1\over8}\Big((p_\nu x^2+x^2p_\nu)B_\mu+B_\mu(p_\nu x^2+x^2p_\nu)\Big),\eqno(A.5)$$
with $A_\mu=Dp_\mu+p_\mu D$ and $B_\mu=Dx_\mu+x_\mu D$ and the $\approx$ symbol means that we are discarding the terms proportional to $\eta_\mn$.

Substituting in (13) gives at order $\alpha^2\beta^4$
$$\eqalignno{&i\Big[[g_{24},x_\mu],p_\nu\Big]+{i\over16}\Big((x_\mu p^2+p^2x_\mu)A_\nu+A_\nu(x_\mu p^2+p^2x_\mu)\Big)-{i\over16}\Big(x_\mu p^2A_\nu+x_\mu A_\nu p^2\cr&+p^2A_\nu x_\mu+A_\nu p^2x_\mu\Big)
\approx i\Big[[g_{24},x_\mu],p_\nu\Big]+{i\over4} p_\mu p_\nu.&(A.6)}$$
The last expression vanishes if
$$[g_{24},x_\mu]=-{i\over8}\Big(Dp_\mu+p_\mu D\Big),\eqno(A.7)$$
and then, up to terms that give contributions proportional to $\eta_\mn$,
$$g_{24}=-{1\over16}(Dp^2+p^2D).\eqno(A.8)$$

At order $\alpha^4\beta^2$, one gets instead
$$\eqalignno{&i\Big[[g_{42},x_\mu],p_\nu\Big]+{i\over16}\Big((p_\nu x^2+x^2p_\nu)B_\mu+B_\mu(p_\nu x^2+x^2p_\nu)\Big)-{i\over16}\Big(p_\nu x^2B_\mu+p_\nu B_\mu x^2\cr&+x^2B_\mu p_\nu+B_\mu x^2p_\nu\Big)
\approx i\Big[[g_{42},x_\mu],p_\nu\Big]+{i\over4} x_\mu x_\nu,&(A.9)}$$
The last expression vanishes up to terms proportional to $\eta_\mn$
if
$$[g_{42},x_\mu]={i\over8}x^2x_\mu,\eqno(A.10)$$
and then
$$g_{42}=-{1\over16}(Dx^2+x^2D).\eqno(A.11)$$

More generally, if we define
$$\hp_\mu(\alpha)=\sum_{m=0}^\infty\alpha^{2m}p_\mu^{(2m)},\qquad\hx_\mu(\beta)=\sum_{n=0}^\infty\beta^{2n}x_\mu^{(2n)},\eqno(A.12)$$
and
$$\hx_\mu=e^{iG}\hx_\mu(\b)e^{-iG}=\hx_\mu(\beta)+\sum_{n=1}^\infty{1\over n!}(\ad_{\,iG})^n\hx_\mu(\beta),\eqno(A.13)$$
with
$$G=\sum_{m,n=1}^\infty\alpha^{2m}\beta^{2n}g_{2m,2n},\qquad h=\sum_{m,n=0}^\infty\alpha^{2m}\beta^{2n}h_{2m,2n},\eqno(A.14)$$
and $h_{0,0}=1$, then from  $[\hx_\mu,\hp_\nu(\alpha)]=i\eta_\mn h$, we get at order $\alpha^{2m}\beta^{2n}$,
$$\eqalignno{&[x_\mu^{(2n)},p_\nu^{(2m)}]+\sum_{{m_1+m_2=m\atop n_1+n_2=n}}\left[\left[ig_{2m_1,2n_1},x_\mu^{(2n_2)}\right],p_\nu^{(2m_2)}\right]\cr
&\ +{1\over2!}\ \sum_{{m_1+m_2+m_3=m\atop n_1+n_2+n_3=n}}\left[\left[ig_{2m_1,2n_1},\left[ig_{2m_2,2n_2},x_\mu^{(2n_3)}\right]\right],p_\nu^{(2m_3)}\right]+\dots=i\eta_\mn h_{2m,2n}&(A.15)}$$
The last term on the left hand side has the form ${1\over k!}\left[\left(\ad_{\,iG_{2,2}}\right)^k(x_\mu),p_\nu\right]$, where $k={\rm min}\,(m,n)$. These relations
can be solved recursively to compute $g_{2m,2n}$ and $h_{2m,2n}$ using the results for $g$ at lower orders.
\end